\documentclass[useAMS,usenatbib]{mn2e}
\usepackage{graphicx}
\title[Vertical shear instability and vortices]{Vortex formation in protoplanetary 
discs induced by the vertical shear instability}
\author[S.Richard, R.P.Nelson \& O.M.Umurhan]{Samuel Richard$^{1}$\thanks{E-mail:
samuel.richard@qmul.ac.uk; r.p.nelson@qmul.ac.uk; orkan.m.umurhan@nasa.gov}, 
Richard P. Nelson$^{1}$ \& Orkan M. Umurhan$^{2,3}$ \\
$^{1}$ Astronomy Unit, Queen Mary University of London, Mile End Rd, London, E1 4NS, U.K.\\
$^{2}$ Space Sciences Division, NASA Ames Research Center, Moffett Field, CA 94035, USA \\
$^{3}$ SETI Institute, 189 Bernardo Way, Mountain View, CA 94043, USA}

\begin{document}

\date{}

\pagerange{\pageref{firstpage}--\pageref{lastpage}} \pubyear{}

\maketitle

\label{firstpage}

\begin{abstract}
We present the results of 2D and 3D hydrodynamic simulations of idealized protoplanetary 
discs that examine the formation and evolution of vortices by the vertical shear 
instability (VSI). In agreement with recent work, we find that discs with radially
decreasing temperature profiles and short thermal relaxation time-scales, are subject to 
the axisymmetric VSI. In three dimensions, the resulting velocity perturbations give 
rise to quasi-axisymmetric potential vorticity perturbations that break-up into discrete 
vortices, in a manner that is reminiscent of the Rossby wave instability. 
Discs with very short thermal evolution time-scales (i.e. $\tau \le 0.1$ local orbit periods) 
develop strong vorticity perturbations that roll up into vortices that have small aspect ratios 
($\chi \le 2$) and short lifetimes ($\sim$ a few orbits). Longer thermal time-scales give 
rise to vortices with larger aspect ratios ($6 \le \chi \le 10$), and lifetimes that depend 
on the entropy gradient. A steeply decreasing entropy profile leads to
vortex lifetimes that exceed the simulation run times of hundreds of orbital periods.
Vortex lifetimes in discs with positive or weakly decreasing entropy profiles are much 
shorter, being 10s of orbits at most, suggesting that the subcritical baroclinic instability 
plays an important role in sustaining vortices against destruction through the
elliptical instability. Applied to the outer regions of protoplanetary discs,
where the VSI is most likely to occur, our results suggest that vortices formed by 
the VSI are likely to be short lived structures.
\end{abstract}


\section{Introduction}

The presence of long-lived vortices in protoplanetary discs has long been considered
as a means of enhancing the planet building process because of the ability
of these anticyclonic structures to capture and concentrate dust grains
\citep{Weizsacker1944,BargeSommeria1995,Johansen2004}.
The presence of vortices in discs can also lead to significant transport of
angular momentum through the excitation of spiral density waves \citep{Johnson2005b}.
In spite of the potential importance of vortices for the evolution of protoplanetary discs
and planet formation, however, the following questions do not yet have definitive answers:
If vortices exist in protoplanetary discs, what are their formation mechanisms?;
Which regions of protoplanetary discs can support the existence of vortices?;
What are the lifetimes of vortices in protoplanetary discs?

A number of hydrodynamic instabilities have been suggested as vortex formation
mechanisms. The Rossby wave instability \citep[RWI;][]{Lovelace99,Li2000,Li2001} has been shown to
produce large scale vortices from an initial axisymmetric state when { {a
non self-gravitating disc has a sufficiently strong local
minimum in the potential vorticity \citep{Umurhan2010,Lovelace2013,Yellin-Bergovoy_etal_2015}}}. 
The extremum can be due to a local maximum in
surface density, such as can appear at the edge of the dead zone where a sharp
change in effective disc viscosity arises \citep[e.g.][]{Varniere2006}, or because
a planet opens a gap in the disc \citep{devalborro2007}.
Another way to produce vortices is through baroclinic instability.
\citet{Klahr2003} introduced the concept of the global baroclinic instability (GBI),
and suggested that a disc with a global negative entropy gradient could form vortices.
More recently, \citet{Petersen2007a} and \citet{Petersen2007b} have shown that vortices can form when 
the disc has an unstable radial stratification and undergoes thermal relaxation (or `cooling').
\citet{Lesur2010} showed in their study that this instability is actually a nonlinear 
instability (requiring finite amplitude perturbations to be activated) and called it the 
subcritical baroclinic instability (SBI). The nonlinearity of the instability means that
the finite amplitude perturbations need to be generated by another unspecified process.
A linear instability that has been suggested as a possible source for these perturbations
is the `convective overstability', in which the growth of epicyclic oscillations is 
powered by the same unstable stratification and thermal relaxation required for the 
subcritical instability, leading to the formation of long-lived vortices \citep{KlahrHubbard2014,
Lyra2014}. Finally, there have been recent suggestions that a zombie vortex instability
may arise in stably stratified flows by the formation of a critical layer 
which rolls up into vortices, which then excite new critical layers and
vortices, leading eventually to space filling turbulence that is dominated
by large vortices \citep{Marcus2013b, Marcus2015}.

In this paper, we examine the possibility that vortices may be formed by
the vertical shear instability (VSI). This is a linear instability that was first
studied by \citet{GoldreichSchubert1967} and \citet{Fricke1968} in the
context of differentially rotating stars. The presence of vertical shear
in protoplanetary discs must arise when there is a radial temperature
gradient, and this can lead to the destabilization of inertial-gravity waves
(oscillations for which rotation and buoyancy provide restoring forces)
when thermal time-scales are shorter than viscous time-scales, and
are of the order of, or shorter than, dynamical time-scales \citep{UrpinBrandenburg1998,
Urpin2003, Nelson2013, BarkerLatter2015, Umurhan2015}. The recent 
study by \citet{Nelson2013} adopted simple
equations of state and cooling prescriptions, and showed that this 
instability can lead to sustained hydrodynamic turbulence with a Shakura Sunyaev 
angular momentum transport parameter $\alpha \sim10^{-3}$ for very short cooling times.
\citet{StollKley2014} performed non-linear hydrodynamic simulations with radiation transport
and showed that the instability operates in the presence of a more complete description of the
gas thermal evolution, albeit with a reduced efficiency of
angular momentum transport. The requirement for very short cooling times suggests 
that this instability is most likely to operate in the outer regions of protoplanetary discs beyond
$\sim 10$ au \citep{ Nelson2013,Umurhan2013}.
An analysis of linear growth rates in discs with energy transport in both the
optically thick and thin regimes presented by \citet{LinYoudin2015} suggests
that the VSI should operate at radii in the range 10 50 au.

This paper is organized as follows. In Sect.~\ref{sec:theory}, we review 
the different processes and instabilities that can lead to the formation
and destruction of vortices in discs, and in Sect.~\ref{sec:numerics} 
we describe the disc models that are the basis of our study and the
numerical scheme used in the simulations. The results of two-dimensional, 
axisymmetric simulations are presented in Sect.~\ref{sec:2D}, and
the results of three-dimensional runs that examine the formation and evolution
of vortices are presented in Sect.~\ref{sec:3D}. Finally we discuss our results
and draw conclusions in Sect.~\ref{sec:discussion}

\section{Theoretical background and expectations}
\label{sec:theory}
Before presenting our simulation results we discuss a number of
theoretical results that are of relevance to this numerical study 
of the VSI. We make use of both spherical
($r$, $\theta$, $\phi$) and cylindrical ($R$, $\phi$, $Z$) coordinates
in this paper. 
The cylindrical coordinates are used in the formulae giving 
the disc structure (density, temperature and velocity profile) for 
convenience, while the spherical coordinates are used for the simulations 
because they fit better with the shapes of the disc models.

\subsection{RWI}
\label{sect:RWI}
The first vortex-forming instability that has been studied in the context of 
protoplanetary discs with near-Keplerian rotation profiles is the RWI. Using a combination of linear analysis and nonlinear 
numerical simulations, \citet{Lovelace99}, \citet{Li2000} and \citet{Li2001} 
showed that a non-axisymmetric instability may develop, leading to the formation of
a number of vortices, when the disc contains a local extremum in the function:
\begin{equation}
\mathcal{L}={\Sigma \over \omega_z}{\left(P \over \Sigma^{\gamma}\right)}^{2/\gamma},
\label{eqn:RWI}
\end{equation}
where $\omega_z$ is the vertical component of the vorticity,
$\Sigma$ is the surface density and $P$ is the pressure
($\omega_z/\Sigma$ is the potential vorticity and $P/\Sigma^\gamma$ is related to the entropy).
It has been observed that these vortices often merge to form a single vortex during 
the advanced stages of nonlinear evolution. Considered within the context of
protoplanetary discs, the RWI is normally observed to develop when 
a local pressure or density maximum is present in the disc, such as may
occur at the edge of a planet-induced gap \citep{deVal-Borro2006, devalborro2007} or at the
edge of a dead zone where there is a sharp transition in the disc viscosity 
\citep{Lyra2012}. Vortices have also been observed to develop spontaneously in
global magnetized disc models that sustain the magnetorotational instability
\citep{FromangNelson2005}. Although this latter phenomenon has not been explored
in detail, a possible explanation is that the vortices arise because the RWI
feeds off the so-called zonal flows that are observed to arise in discs with
MHD turbulence \citep{Steinacker2002, Papaloizou2003, Johansen2009, Bai2014}.

\subsection{SBI}
\label{sect:SBI}
The existence of a vortex-forming baroclinic instability operating
in protoplanetary discs was first suggested by \citet{Klahr2003},
based on a series of nonlinear simulations conducted using disc models
with negative radial entropy gradients. The linear properties of this
GBI were investigated by \citet{Klahr2004}
and \citet{Johnson2005b}, who found evidence for only transient growth.
The nonlinear evolution was investigated using shearing box simulations
by \citet{JohnsonGammie2006}, who found no instability. 
\citet{Petersen2007a} examined disc stability using global anelastic 
simulations of baroclinic discs with prescribed cooling and observed the
growth of vortices that were found to survive for hundreds of orbits
\citep{Petersen2007b}. In a subsequent paper, \citet{Lesur2010}
used both incompressible and compressible shearing box simulations
to examine the growth and survival of vortices in discs with radial
entropy gradients and imposed cooling, finding that these discs are
unstable to a finite-amplitude instability that leads to the formation
of long-lived vortices -- the SBI.

The SBI is a non-linear convective instability that leads to the formation
and amplification of vortices when the radial stratification satisfies
the Schwarschild instability criterion:
\begin{equation}
N_R^2<0
\end{equation}
and when the flow undergoes thermal relaxation.
$N_R$ is the radial Brunt Vaisala frequency and is defined by
\begin{equation}
N_R^2=-{1 \over C_{\rm p} \rho}{\partial P \over \partial R}{\partial S \over \partial R}
\end{equation}
where $\rho$ is the density, $P$ is the pressure, $S=C_{\rm p} \ln(P^{1/\gamma}\rho^{-1})$ 
is the entropy per unit mass, and $C_{\rm p}$ is the specific heat capacity at 
constant pressure.
The thermal relaxation time scale must be of the order of one orbital period. Too short a
time scale prevents vortex formation, while too long a time scale does not allow
vortices to be amplified.
When the stratification is stable ($N_R^2>0$) vortices are observed to form,
but instead of being amplified they decay and dissolve into the background flow.
The non-linear character of this instability implies that sufficiently strong 
perturbations are required to trigger it. 

\subsection{Convective overstability}
\label{sect:convective}
The convective overstability is a linear instability that depends on having 
$N_R^2 < 0$ and thermal relaxation on $\sim$ dynamical time-scales
\citep{KlahrHubbard2014, Lyra2014}. The instability involves the growth
of horizontal epicyclic oscillations, hence the name by which it is known,
and according to the study by \citet{Lyra2014} it leads to the formation of 
long lived vortices. It has been suggested as a possible source of the finite
amplitude perturbations required for the SBI to operate.
Vertical buoyancy has not been included in the analysis of this instability so far, 
so its behaviour in vertically stratified discs remains unexplored. 

\subsection{VSI}
The Rayleigh criterion for hydrodynamic stability is satisfied by discs with
strictly Keplerian rotation profiles because
\begin{equation}
\frac{d j^2}{dR} > 0,
\end{equation}
where $j=R^2 \Omega(R)$ is the specific angular momentum.
More generally, a disc with angular velocity varying with both radius and height, 
$\Omega(R,Z)$, that is subject to adiabatic, axisymmetric perturbations is stable 
according the Solberg Hoiland criteria \citep[e.g.][]{Tassoul78}. For accretion discs 
with negative radial and vertical pressure gradients these stability criteria can 
be written
\begin{equation}
\frac{1}{R^3} \frac{\partial j^2}{\partial R} + \frac{1}{\rho C_{\rm p}}
\left(\left| \frac{\partial P}{\partial R} \right| \frac{\partial S}{\partial R}
+ \left| \frac{\partial P}{\partial Z} \right| \frac{\partial S}{\partial Z} \right) > 0
\label{eqn:SH1}
\end{equation}
\begin{equation}
\frac{\partial j^2}{\partial R} \frac{\partial S}{\partial Z} - 
\frac{\partial j^2}{\partial Z}\frac{\partial S}{\partial R} > 0.
\label{eqn:SH2}
\end{equation}
For a nearly inviscid disc in which perturbations are no longer adiabatic,
the thermal evolution of perturbed fluid elements can remove the stabilizing influences
of entropy gradients when the cooling time is short enough. This leads to the
well-known Goldreich Schubert Fricke instability \citep{GoldreichSchubert1967, Fricke1968},
which in the context of accretion discs is known as the VSI \citep{Urpin2003}.
The VSI develops when the flow is vertically sheared and almost locally isothermal.
The instability criterion is:
\begin{equation}
{\partial j^2 \over \partial R} -{k_R \over k_Z}{\partial j^2 \over \partial Z}<0.
\end{equation}

Vertical shear is always present in a protoplanetary disc, unless the flow is isothermal 
or homentropic, and $|\partial j^2 / \partial R| \gg |\partial j^2 / \partial Z|$, so 
the instable modes have $k_R \gg k_Z$ (radial wavelengths are much shorter than vertical 
wavelengths). In the locally isothermal limit, the maximum growth rate of the VSI depends 
on the temperature profile and the scaleheight :
\begin{equation}
\Gamma_{max} \sim |q| \left( H \over R \right) \Omega,
\end{equation}
where $q$ is the temperature profile power-law index. In a recent study,
\citet{Nelson2013} showed that the VSI can cause a disc to
become highly turbulent in the locally isothermal regime, with velocity
perturbations having very short radial wavelengths such that there are
strong local gradients in the flow. The only simulations conducted in 3D in
that work utilized a locally isothermal equation of state, and the resulting
turbulence led to the excitation of spiral density waves in the flow but no
obvious signs of long lived vortices. It therefore remains an open question
whether or not the VSI can lead to the formation of perturbations that generate
vortices when thermal relaxation is not treated as being instantaneous, perhaps 
through the RWI or the SBI acting on the primary perturbations generated by the VSI.

\subsection{Elliptical instability}
\label{sect:Elliptical}

\citet{Lesur2009a}, using a local approach, showed that 3D elliptical vortices 
may be unstable. The vertical modes ($\vec{k}=k_z\vec{e_z}$) are the dominant 
modes in vortices with an aspect ratio $1.5<\chi<4$, and have a growth rate 
given by:

\begin{equation}
\Gamma={\cal S} \sqrt{-\left({{2 \Omega} \over {\cal S}}- 
{\chi \over {\chi-1}}\right)\left({{2 \Omega} \over {\cal S}}- 
{1 \over {\chi(\chi-1)}}\right)}
\end{equation}

where ${\cal S}$ is the shear and $\chi$ the aspect ratio of the vortex. 
In the case of a Keplerian protoplanetary disc ${\cal S}=1.5\Omega$, 
and the growth rate is:
\begin{equation}
\Gamma={3 \over 2} \Omega \sqrt{-\left({4 \over 3}- 
{\chi \over {\chi-1}}\right)\left({4 \over 3}- {1 \over {\chi(\chi-1)}}\right)}
\end{equation}

The fact that the mode is purely vertical and the growth rate is independent of
the wavelength makes the instability quite easy to capture and 
very high resolution simulations are not required for it to be resolved. 
It should be noted that this result,
however, is valid in the local approximation and is not true for long wavelengths.

The case of larger aspect ratio vortices is more complex because the instability 
is fully three dimensional. The instability is due to the resonance between inertial 
waves in an unstratified flow or high frequency buoyancy waves in a stratified flow 
and the turnover frequency of the vortex.
As no inertial modes can match the turnover frequency in a vortex with $4<\chi<5.9$, 
these vortices are stable when the flow is not stratified, while in a stratified flow 
vortices are always unstable.

However, despite the unstable character of vortices in that case, the elliptical 
instability is difficult to observe because of the high resolution needed to resolve it.
The unstable mode will have $k_{\phi}^{max} \sim k_z$ and $k_r^{max} \sim \chi k_z$, 
so the radial resolution needed to resolve it depends on the aspect ratio.
Moreover, the growth rate for $\chi>6$ is about 50 time smaller than for vortices with 
$\chi<4$. These two points make the elliptical instability difficult to observe in 
numerical simulations that contain large aspect ratio vortices.

\subsection{Streaming instability}
\label{sect:Streaming}

Although the streaming instability applies to a disc composed of 
interpenetrating gas and solids, rather than to the single fluid system
considered here, we discuss it briefly for completeness. It arises as 
a linear instability when aerodynamic drag causes inwards radial drift of 
solid particles, and the backreaction on the gas is included in the 
dynamics \citep{YoudinGoodman2005}. The linearly growing modes 
consist of particle density enhancements with growth times that 
lie between fast dynamical time-scales and slower radial drift time-scales,
and maximal growth rates arise for particle stopping times comparable
to dynamical time-scales and when the local solids-to-gas ratio is
of the order of unity. Nonlinear simulations indicate that it can lead to
severe clumping of solids, such that particle concentrations are able
to collapse directly to form planetesimals \citep{Johansen2007}.
As such, this provides an alternative to the concentration of dust in
vortices as a means of forming planetesimals in protoplanetary discs.

\section{Numerical method}
\label{sec:numerics}
The equations of motion are for a compressible and inviscid fluid subject to
a central gravitational potential:
\begin{equation}
{\partial \rho \over \partial t} + \vec{\nabla}.\left(\rho \vec{V}\right)=0,
\end{equation}

\begin{equation}
{\partial \rho \vec{V} \over \partial t} + \vec{\nabla}.\left(\rho \vec{V} \vec{V}\right)
 =-\vec{\nabla} P - \rho \vec{\nabla}\Phi,
\end{equation}

\begin{equation}
{\partial \rho E \over \partial t} + \vec{\nabla}.\vec{V}(\rho E+P) =\rho \vec{V} .\vec{\nabla}\Phi 
- {\rho \over \gamma-1} {T-T_0 \over \tau}.
\label{eqn:Energy}
\end{equation}
Here $\rho$ is the density, $\vec{V}$ the velocity, $P$ the pressure and $E$ the 
total energy per unit mass, and 
$\Phi=-GM/r$ the gravitational potential due to the central star.
The system is closed using the perfect gas equation of state: 
$\rho E=P/(\gamma-1)+1/2\rho \vec{V}^2$, where we adopt $\gamma=7/5$.
The last term in the energy equation (\ref{eqn:Energy}) is a thermal relaxation term: 
the temperature relaxes to the initial temperature $T_0$ with a relaxation time $\tau$.
$\tau$ is assumed to be a function of $R$, the cylindrical radius, being a fixed multiple 
or fraction of the local Keplerian orbital period.

We adopt a radial power law for the disc temperature while assuming
that the disc is initially isothermal in the vertical direction,
and we also adopt a radial law for the midplane density:
\begin{eqnarray}
T(R) &= & T_0 \left( {R \over R_0}\right)^q \nonumber \\
\rho(R,0) & = & \rho_0 \left( \frac{R}{R_0}\right)^p,
\end{eqnarray}
where $R_0$ is a reference radius.
Other quantities of interest are determined using the equations of force balance in the
radial and vertical directions:
\begin{equation}
{\partial P \over \partial R} = -{GMR\over r^3}+R\Omega^2 
\end{equation}
\begin{equation}
{\partial P \over \partial Z} = -{GMZ\over r^3}
\end{equation}
where $r=\sqrt{R^2+Z^2}$ is the spherical radius. 

The equilibrium solutions for density and angular velocity give
\begin{equation}
\rho(R,Z)=\rho_0\left({R\over R_0}\right)^p \exp\left({GM \over c_s^2}\left[{1 \over r}-{1 \over R}\right]\right) 
\end{equation}
\begin{equation}
\Omega(R,Z)=\Omega_k\sqrt{1+(p+q)\left(H\over R\right)^2+q\left(1-{R\over r}\right)}
\end{equation}
where $\Omega_k=\sqrt{GM/r}$ is the Keplerian velocity and $H$ is the 
scaleheight defined through:
\begin{equation}
H={c_s \over \Omega_k}.
\end{equation}
$c_s$ is the isothermal sound speed defined through:
\begin{equation}
c_s^2 ={P\over \rho},
\end{equation}
and we define the local disc aspect ratio $h=H/R$.

In the disc model considered here, the Brunt-Vaisala frequency in the midplane is:
\begin{equation}
N_R^2={T  \over \gamma R^2}(p+q)(p(\gamma-1)-q)
\end{equation}
then the radial stratification is stable when $q<p(\gamma-1)$ and unstable otherwise.

These equations are solved in spherical coordinates $(r,\theta,\phi)$
using a finite volume code using the MUSCL Hancok method \citep{Richard2013}
We choose reflecting boundary conditions at the radial and vertical
boundaries because of their ease of implementation, and because 
\citet{Nelson2013} have shown that the boundary conditions appear to have no 
effect on the development of the VSI.

The computational domains of our simulations are as follows. In radius 
the inner boundary is located at $r=1$ and the outer boundary at $r=1.5$,
and the azimuthal domain runs between $\phi=0$ and $\phi=\pi/4$.
The meridional domain extends $\pm 5 \times h$ above and below the disc midplane.
In all simulations except those with $h=0.05$, we use 500 grid cells in the radial direction.
For models with $h=0.05$, we double the radial resolution and use 1000 grid cells.
All runs use 300 grid cells in the azimuthal direction and 200 cells in the meridional
direction.
The equilibrium state is perturbed by adding $10^{-6}c_s$ amplitude white 
noise to each component of the velocity in all simulations presented in 
this paper.

\section[]{Axisymmetric simulations}
\label{sec:2D}

An investigation of the potential for the VSI to generate vortices in discs
clearly requires 3D simulations to be performed. The fact that the unstable
VSI modes have radial wavelengths much shorter than vertical scales indicates
that high resolution simulations are required, leading inevitably to long 
simulation run times. In light of this, most of the simulations presented
in this paper adopted density and temperature power-law profiles $p=-1.5$
and $q=-2$, where the adoption of the steep temperature profile reduces
the growth time scale for the VSI, hence allowing a suite of 3D simulations
to be undertaken. We note that these disc models are stable according to
the Solberg-Hoiland criteria given by equations (\ref{eqn:SH1}) and (\ref{eqn:SH2})
and have imaginary values of the radial Brunt-Vaisala frequency $N_R$.
Depending on the adopted thermal relaxation time scale, these models 
may be unstable to the SBI and the convective overstability. It is
therefore possible that any vorticity perturbations generated by the
VSI may be amplified and sustained by the SBI, leading to long lived 
vortices.

Given that the parameters used for the simulations in this paper were not considered
in the study presented by \citet{Nelson2013}, we have undertaken a suite of 2D 
axisymmetric simulations to examine the growth times of the VSI in these models,
and to also examine the critical cooling times that allow the VSI to operate.
We consider models with $h=0.2$, 0.1 and 0.05 with different thermal relaxation 
times.

Following \citet{Nelson2013}, the total kinetic energy is defined by the volume 
integral of the sum of the radial and meridional kinetic energies, normalized by the 
volume integral of the initial azimuthal kinetic energy:

\begin{equation}
Ec={\int_V \rho (v_{\theta}^2 +v_{r}^2)dV \over \int_V \rho v_{\phi_0}^2dV}
\end{equation}

\begin{figure}
\includegraphics[width=84mm]{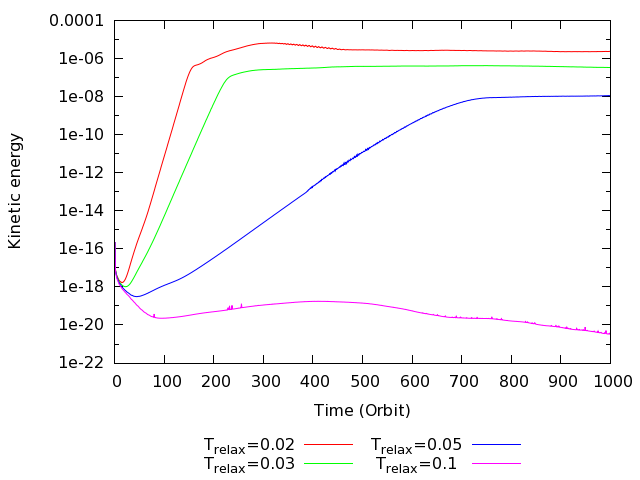}
\caption{Kinetic energy for a disc with a $h/r=0.05$ and different values of the cooling time.}
\label{2Dh=0.05}
\end{figure}

\begin{figure}
\includegraphics[width=84mm]{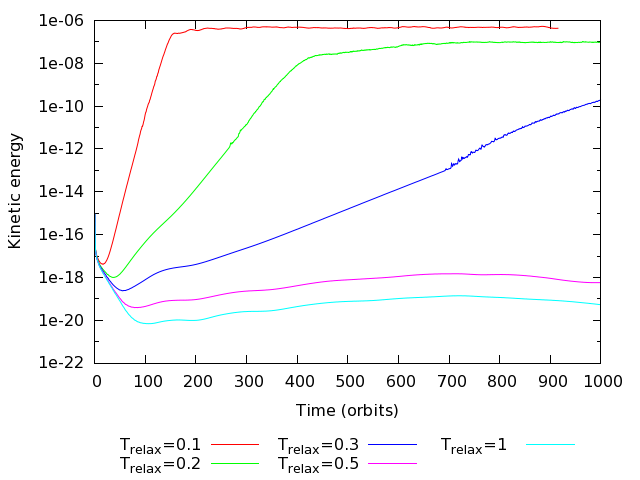}
\caption{Kinetic energy for a disc with a $h/r=0.1$ and different values of the cooling time.}
\label{2Dh=0.1}
\end{figure}

\begin{figure}
\includegraphics[width=84mm]{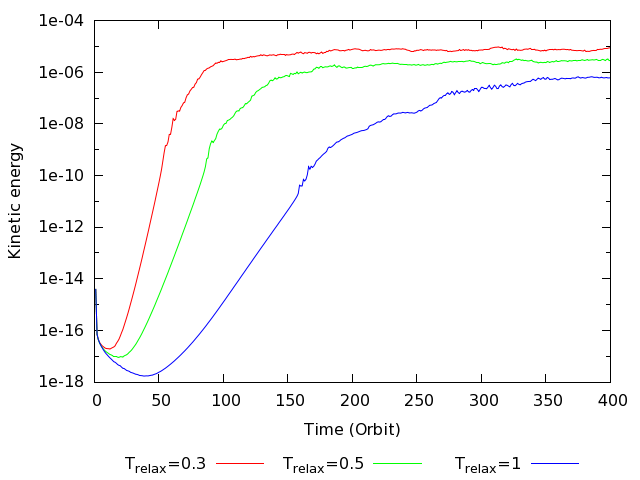}
\caption{Kinetic energy for a disc with a $h/r=0.2$ and different values of the cooling time.}
\label{2Dh=0.2}
\end{figure}

\begin{figure*}
\includegraphics[height=50mm]{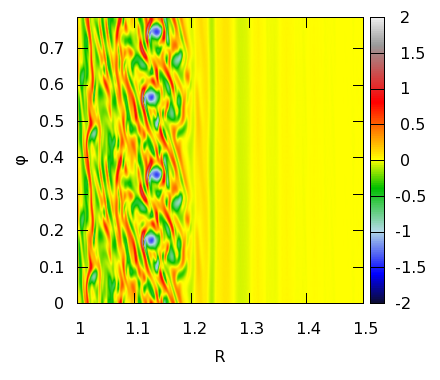}
\includegraphics[height=50mm]{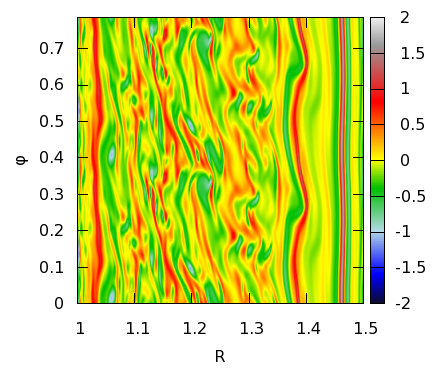}
\includegraphics[height=50mm]{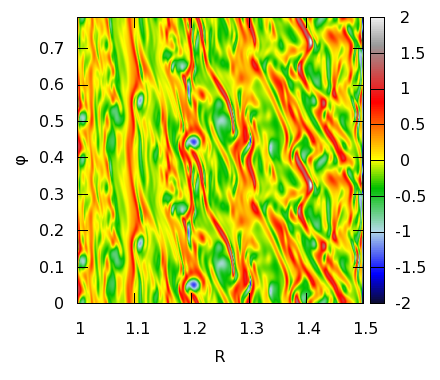}
\includegraphics[height=70mm]{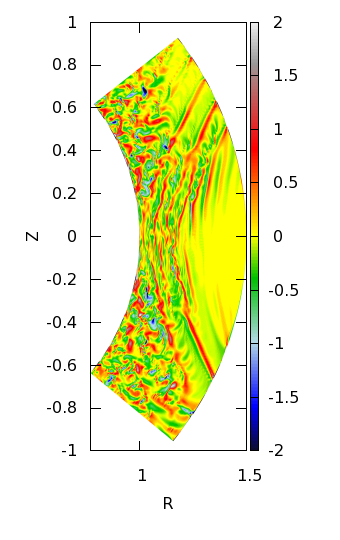}
\hspace{10mm}
\includegraphics[height=70mm]{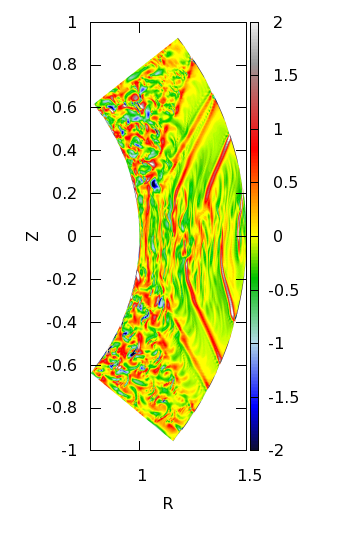}
\hspace{10mm}
\includegraphics[height=70mm]{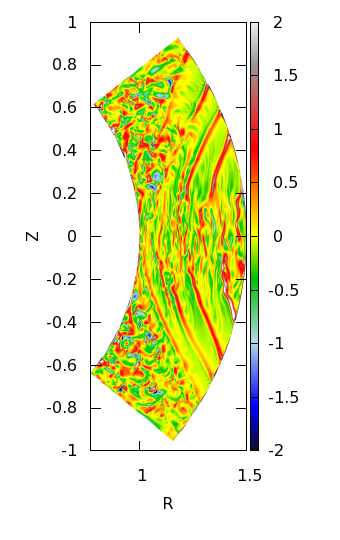}

\caption{Vorticity profile in the midplane and meridional plane after 34, 54 and 64 orbital periods, 
with $H/R=0.2$ and $\tau=0.1$.}
\label{tau=0.1}
\end{figure*}

The results of the 2D simulation are shown in Figs. \ref{2Dh=0.05}, \ref{2Dh=0.1} and \ref{2Dh=0.2},
which display the evolution of the total kinetic energy in the disc models with $h=0.05$, $0.1$ 
and $0.2$ respectively. For each value of $h$ the growth rate decreases when the cooling time 
increases, in agreement with \citet{Nelson2013}, and for a critical value of the cooling time
the disc becomes stable to the VSI. This critical value depends strongly on the scaleheight. 
For $h=0.05$, the disc become stable for a cooling time $0.05<\tau<0.1$, for $h=0.1$ it stabilizes
for $0.3<\tau<0.5$, and for $h=0.2$ the disc remains unstable for a cooling time as large as one 
orbital period.

We notice that the level of saturation also depends on the cooling time: the total kinetic 
energy during the saturated state is higher for a shorter cooling time, and decreases for 
longer cooling times. In other words, higher amplitude velocity and vorticity perturbations
are expected for shorter cooling times when we consider 3D simulations below.

\section{3D simulations}
\label{sec:3D}

\subsection{Fiducial model}
The primary goal of this paper is to examine whether or not the nonlinear development of the
VSI leads to the formation of vortices. Secondary goals include determining the range of
conditions under which vortices form, understanding the nature of these vortices as a 
function of system parameters, and the possible roles of the RWI and the SBI in creating
and maintaining vortices. Given our interest in the potential role of the SBI, we have
chosen a disc model which in principle allows the development of the VSI and SBI.
Both instabilities are more efficient in thick discs, so we choose $h=0.2$. 
Our 2D runs described in Section~\ref{sec:2D} indicate that this model remains unstable to the 
VSI even for relatively long cooling times, which are necessary for the SBI to operate 
\citep{Lesur2010}.

Our fiducial model has a thermal relaxation time $\tau=0.1$ local orbits. The 2D simulations
described in Section~\ref{sec:2D} indicate that the growth rate of the VSI should be quite large
in this case, and that this cooling time is significantly shorter than the critical value 
for which the VSI no longer operates.
During the early phases the VSI in 3D remains axisymmetric and so develops very similarly 
to 2D simulations \citep{Nelson2013}. The axisymmetric velocity perturbations correspond
naturally to axisymmetric vorticity perturbations which grow with time. When these axisymmetric 
vorticity bands reach a critical amplitude, they tend to destabilize and vortices are formed.
This evolution is illustrated by Fig.~\ref{tau=0.1} which shows contours of the perturbed
vertical component of the vorticity in the midplane (top panels) and in a slice along
the meridional plane (bottom panels). As discussed in \citet{Nelson2013}, the VSI is
first observed at high latitudes in the disc and descends down towards the midplane,
as seen in Fig.~\ref{tau=0.1}. 

The bottom panels of Fig.~\ref{tau=0.1} also show that the 
break-up of the initially axisymmetric vorticity
perturbations into discrete vortices produces structures with relatively small length scales
in the vertical direction. In other words, the vortices formed in this simulation are not
large scale columnar structures, but instead appear to be coherent over vertical length scales
that are significantly shorter than the vertical scaleheight (i.e $\sim 0.1 H$). 
Although we only plot the vorticity in the disc midplane, we find that vortices form at 
all heights in the disc.

Looking in particular at the first and last of the top panels in Fig.~\ref{tau=0.1}, we see 
that the vortices have relatively small aspect ratios when projected in the $R$-$\phi$ plane. 
Detailed measurements indicate that the 
aspect ratio $\chi \sim 2$ for these vortices. As expected from our discussion of the elliptical
instability presented in Section~\ref{sect:Elliptical}, these vortices do not survive for very long, 
and we measure a typical life time of between 2 and 3 orbits. Inspection of animations of the midplane 
vorticity indicates that this disc model develops a vigorous turbulent flow in which vortices 
continuously appear and disappear on time-scales of a few orbits.

\begin{figure}
\includegraphics[height=50mm]{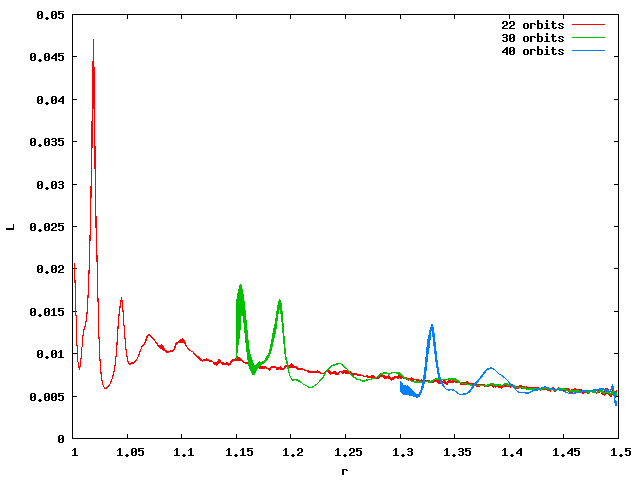}
\caption{Profiles of the quantity ${\mathcal L}$ demonstrating the
development of extrema in this quantity and hence the conditions
required for the RWI to operate and form vortices.}
\label{fig:L-RWI}
\end{figure}

Examination of Fig.~\ref{tau=0.1} indicates that the formation mechanism of the vortices
is the creation of narrow, axisymmetric vorticity perturbations that then break up into discrete
vortices when the perturbation amplitude becomes large enough. 
From our discussion of the
RWI in Section~\ref{sect:RWI}, we can see that perturbations to the vertical component of 
the vorticity that are narrowly confined in radius correspond to the creation of extrema in 
the quantity ${\mathcal L}$ defined by equation~(\ref{eqn:RWI}). As such, it appears 
that vortices form in 
this simulation because the VSI generates narrow vorticity perturbations and extrema in 
${\mathcal L}$ that destabilize through the RWI. The SBI does not appear to play an 
important role in this particular simulation.
\par
{{
Regarding the mechanism of roll-up observed in our simulations, 
the axisymmetric VSI is accompanied by radial
azimuthally symmetric pressure perturbations which
induces jets in the zonal (azimuthal) direction.  Such jets are 
characterized by narrow abutting azimuthal strips of positive/negative vertical vorticity
(recall vorticity anomalies are here understood to be with respect to the background Keplerian
frame).  When the amplitude of the jet gets large enough, the RWI induces 
the roll-up of the negative vorticity strip
leaving the positive vorticity strip more or less intact since 
positive vorticity anomalies in non self-gravitating discs are stable
to the RWI \citep{Umurhan2010}.  
An examination of the top row of Fig.s in Fig.~\ref{tau=0.1} shows exactly
this pattern, where the negative vorticity anomalies created by the VSI eventually roll up into
localized vorticity while leaving the positive vorticity strips alone.
The profile of ${\mathcal L}$ at different 
times during this simulation is shown in Fig.~\ref{fig:L-RWI}, illustrating the development
of the secondary RWI as caused by the VSI throughout many stages
and radial locations of the simulation. }}

\subsection{Dependence on viscosity}

\begin{figure}
\includegraphics[width=84mm]{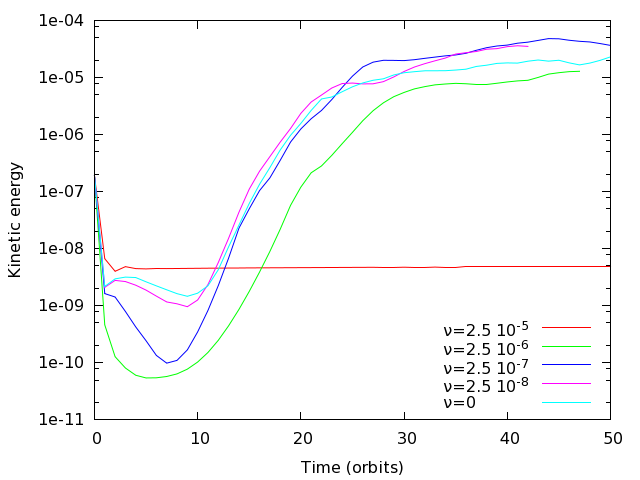}
\caption{Kinetic energy for a disc with a $h/r=0.2$, $\tau=0.1$ and different viscosity.}
\label{viscosity_grow}
\end{figure}

\begin{figure*}
\includegraphics[height=35mm]{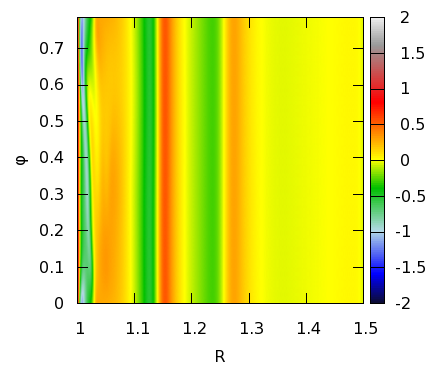}
\includegraphics[height=35mm]{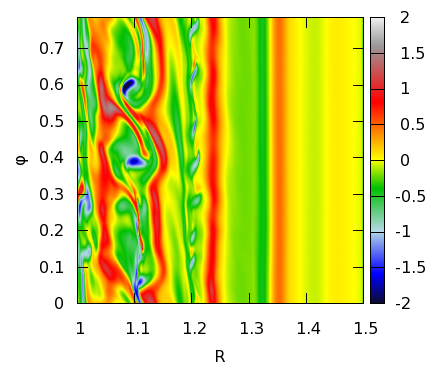}
\includegraphics[height=35mm]{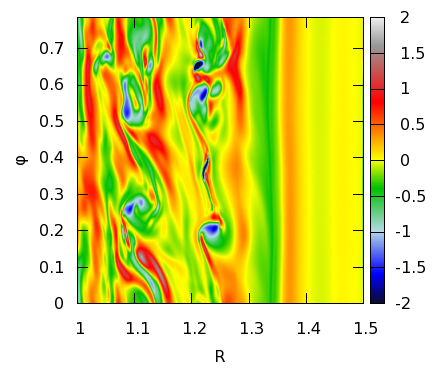}
\includegraphics[height=35mm]{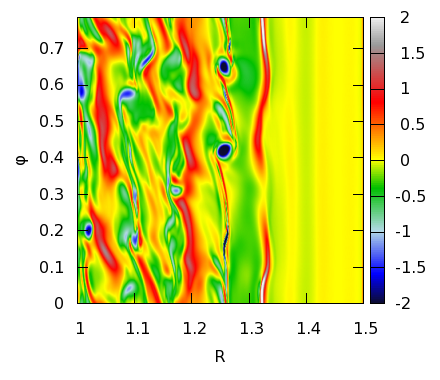}
\caption{Vorticity in the midplane after 26 orbits and for $\nu=2.5~\times~ 10^{-6}$, $2.5~\times~ 10^{-7}$, $2.5 ~\times~10^{-8}$ and $0$.}
\label{viscosity_profile}
\end{figure*}

In this section we investigate the role of viscosity in the 
development of the instability.
We have performed several simulations using the same disc model 
described in the previous section, but adding viscous stresses 
and varying the kinematic viscosity, $\nu$.
Fig. \ref{viscosity_grow} shows the growth of the instability for 
different values of $\nu$.
As expected, it shows that the viscosity has a stabilizing effect 
on the VSI. The growth rate increases when the viscosity decreases 
until $\nu=2.5 \times 10^{-8}$, for which the behaviour is close to the 
inviscid case.
A large viscosity totally inhibits the development of the VSI.
The vorticity contours for the different viscosities are plotted 
in Fig. \ref{viscosity_profile} after simulation run times of
26 orbits. For $\nu=2.5 \times 10^{-6}$ the flow is still axisymmetric 
and the vortices have not yet formed, while for the lower viscosity 
($\nu=2.5 \times 10^{-8}$) the vorticity profile is very similar to the 
inviscid case. This suggests that the numerical diffusion is of
the order of $10^{-8}$. This value correspond to a Reynolds number 
$Re=Hc_s/\nu\approx 10^6$ and a Shakura-Sunyaev viscous stress 
parameter $\alpha=2.5\times 10^{-7}$ \citep{Shakura73}.
This value is too small to be responsible of the short lifetimes of 
the vortices, so we are  confident that the disappearance of the 
vortices has a physical origin rather than a numerical one.

\begin{figure*}
\includegraphics[height=50mm]{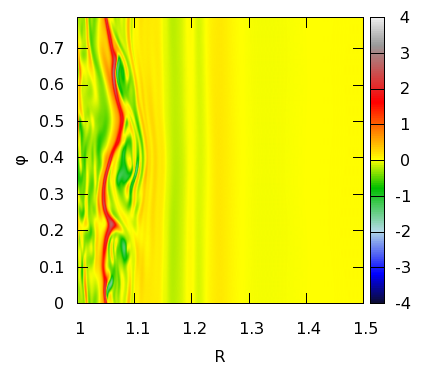}
\includegraphics[height=50mm]{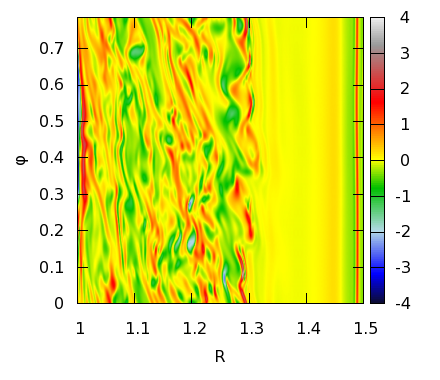}
\includegraphics[height=50mm]{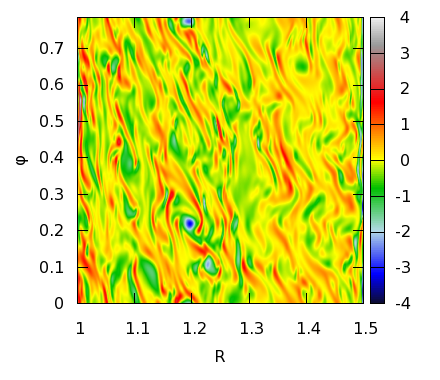}\\
\includegraphics[height=70mm]{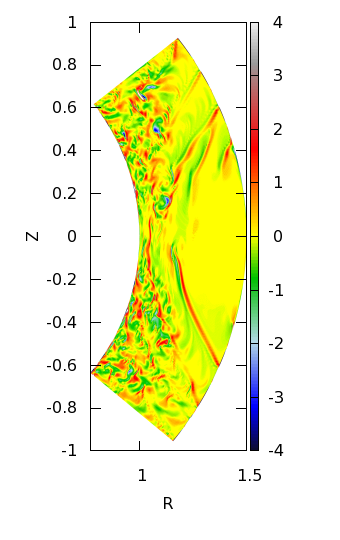}
\hspace{10mm}
\includegraphics[height=70mm]{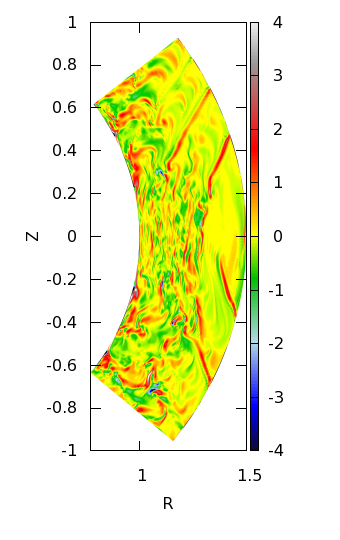}
\hspace{10mm}
\includegraphics[height=70mm]{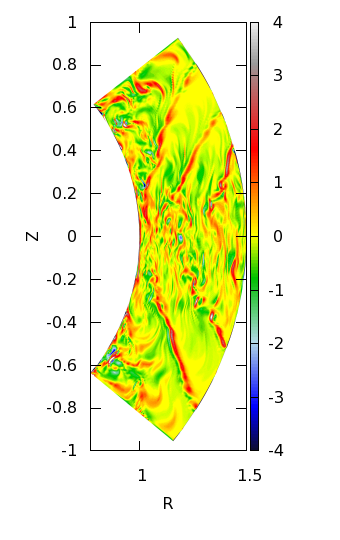}

\caption{Vorticity contours in the midplane and meridional plane after 33, 41 and 74 orbital periods, 
with $h=0.2$ and $\tau=0.05$.}
\label{tau=0.05}
\end{figure*}

\begin{figure*}
\includegraphics[height=50mm]{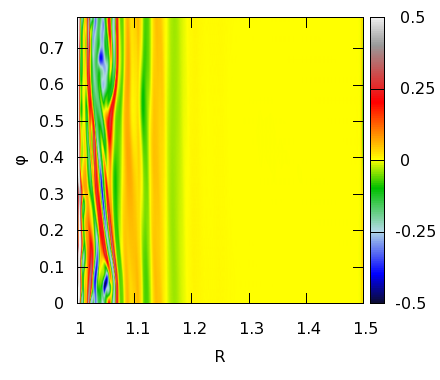}
\includegraphics[height=50mm]{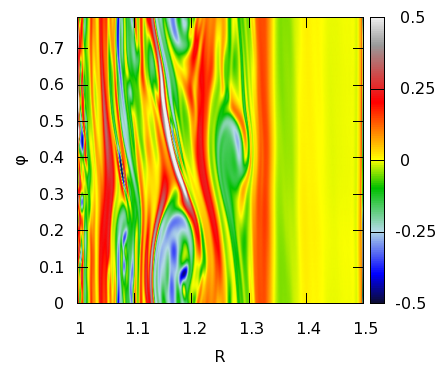}
\includegraphics[height=50mm]{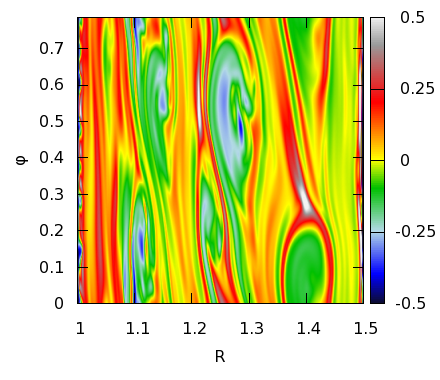}\\
\includegraphics[height=70mm]{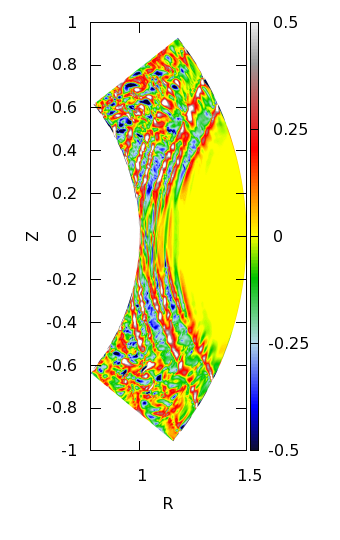}
\hspace{10mm}
\includegraphics[height=70mm]{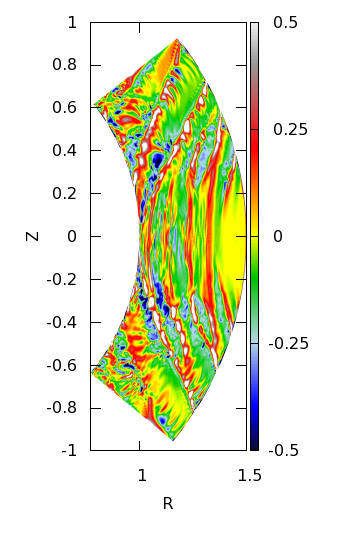}
\hspace{10mm}
\includegraphics[height=70mm]{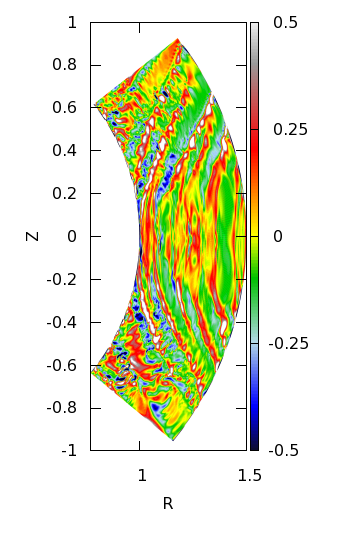}

\caption{Vorticity profile in the midplane and meridional plane after 201, 317 and 401 orbital periods, 
with $h=0.2$ and $\tau=0.5$.}
\label{tau=0.5}
\end{figure*}

\subsection{Dependence on cooling time}
\label{sect:cool-time}
In this section, we investigate how vortex formation depends on the thermal relaxation
by considering the cooling times $\tau=0.05$ and $0.5$, while keeping all
other disc parameters the same. Inspection of Fig.~\ref{2Dh=0.2} shows that
the $\tau=0.05$ run should lead to rapid growth of the VSI, and the generation
of relatively large velocity of vorticity perturbations. A cooling time of
$\tau=0.5$ should lead to a longer growth time for the VSI and weaker velocity and
vorticity perturbations.

Fig. \ref{tau=0.05} shows contours of the vertical vorticity perturbations for the
$\tau=0.05$ run. Comparing the spectrum bars for the contours shown in Fig.~\ref{tau=0.1}
and \ref{tau=0.05} shows that the latter run generates significantly larger vorticity 
perturbations. The consequence of this is that the vortices formed in the nonlinear
saturated state of this simulation have smaller aspect ratios, $\chi$, and shorter
lifetimes. Close inspection of the top panels in Fig.~\ref{tau=0.05} and of
sequences of snapshots similar to these panels indicates that $\chi\sim 1.5$ for
this run, with vortex lifetimes being approximately one orbit.

Fig.~\ref{tau=0.5} show the results of the simulations with $\tau=0.5$.
Inspection of the spectrum bar that indicates the amplitude of the vorticity
perturbations shows that this run produces significantly weaker vorticity 
perturbations than the runs with $\tau=0.1$ and 0.05. 
Consequently, the nature of the flow is very different in this case,
consisting of elongated and long-lived vortices. We estimate that the
typical vortex aspect ratio in this run is $\chi \sim 6$ and the 
average life time exceeds the simulation run times. Close inspection
of the vortices indicates that they maintain a turbulent core
throughout the evolution, presumably due to the elliptical instability 
operating in this case. The long-lived nature of these vortices suggests
that their survival is due to the action of the SBI, that
continuously attempts to increase the amplitude of the vortices in
opposition to the elliptical instability which attempts to destroy them.
Note, however, that this statement is somewhat speculative
since we may not be able to fully resolve the elliptical instability
in this case.

Examining the vertical structure of the vortices in this case, we note that
they occupy a greater height in the disc than those observed in the runs
with $\tau=0.1$ and 0.05. The vortices shown in the upper panels of
Fig.~\ref{tau=0.5} extend above and below the midplane by approximately
one scaleheight. 

In summary, we find that the VSI gives rise to relatively large amplitude 
axisymmetric vorticity perturbations when the cooling time is short 
(i.e. $\tau \le 0.1$ orbits), and this leads to the formation of small 
aspect ratio ($\chi \le 2$) vortices that form through the RWI 
and which extend only a small distance in height (approximately 10\% of
the local scaleheight). The
lifetimes of these vortices is found to be very short, being of the
order of a few orbital periods. A longer cooling time of $\tau=0.5$ orbits 
gives rise to lower amplitude vorticity perturbations, leading to the formation
of elongated (i.e. $\chi \sim 6$) vortices that extend in height by approximately
2 scaleheights and have lifetimes that exceed the
simulation run times. These vortices are observed to have turbulent cores,
presumably due to the elliptical instability, suggesting that their long
lifetimes are due to the action of the SBI maintaining the integrity
of these structures. Table \ref{Tab1} summarize the growth rate of the instability and the vorticity and the aspect ratio of the resulting vortices for each simulations.
 
\begin{table}
\begin{center}
\begin{tabular}{|c|c|c|c|}
\hline
Cooling time & Growth rate      & Vorticity         & Vortex aspect ratio \\
(orbits)     &  (orbits)$^{-1}$ & $\omega_z/\Omega$ &      \\
\hline
0.5 & 0.31 & -0.3 & 6 \\
0.1 & 0.9 & -1.1 & 2 \\
0.05 & 1.7 & -3 & 1.5\\
\hline
\end{tabular}
\caption{\label{Tab1} Results for the $h=0.2$, $q=-2$, $p=-1.5$ simulations.
First column shows cooling time, second column shows growth rate of the VSI,
third column shows peak vorticity
perturbation (in units of the local orbital angular velocity), 
and fourth column shows mean aspect ratio of the emerging vortices.}
\end{center}
\end{table}

\subsection{Dependence on scaleheight}
In the previous simulations the scaleheight was set to $h=0.2$, which corresponds to a rather 
thick disc that may be unrealistic for a typical protoplanetary disc. Here, we investigate the 
evolution of two thinner disc models. As shown in Section \ref{sec:2D}, for thinner discs we need 
shorter cooling times to trigger the VSI. We present a simulation with $h=0.1$ and $\tau=0.2$.
Inspection of the results described in Section~\ref{sec:2D} indicates that the growth time of
the VSI in this case will be quite long as the cooling time is relatively close to the critical
value for which the VSI switches off. Thus, we might reasonably expect behaviour that is similar 
to that shown by the run with $h=0.2$ and $\tau=0.5$ in this case. We also present a simulation
with $h=0.05$ and $\tau=0.01$. The results presented in Section~\ref{sec:2D} indicate that this
model should experience rapid growth of the VSI as the cooling time is much shorter than the
critical thermal relaxation time scale. We might reasonably expect results similar to those
observed for the runs with $h=0.2$ and $\tau=0.1$ and 0.05 in this case. In order to resolve
the smaller length scale features expected in this run the simulation was conducted with
double the resolution in the radial direction.

\begin{figure*}
\includegraphics[height=50mm]{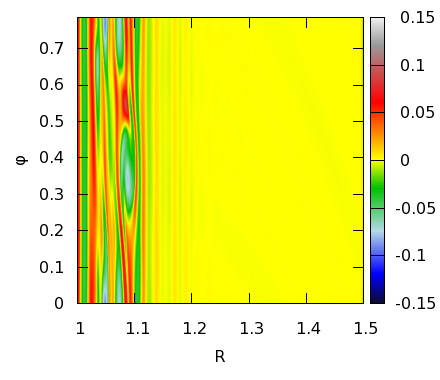}
\includegraphics[height=50mm]{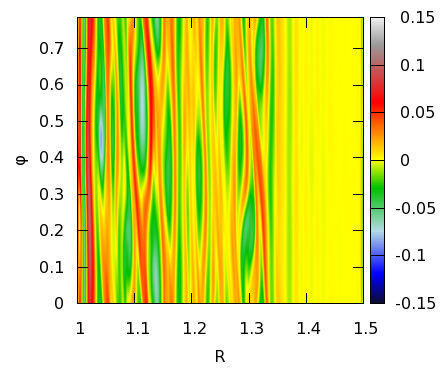}
\includegraphics[height=50mm]{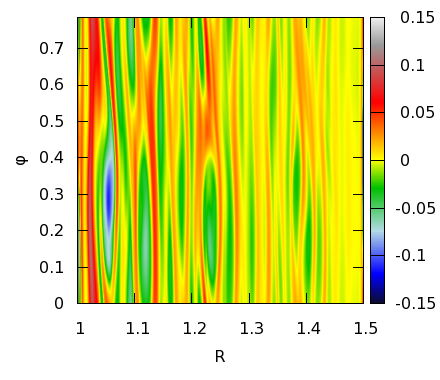}\\
\includegraphics[height=65mm]{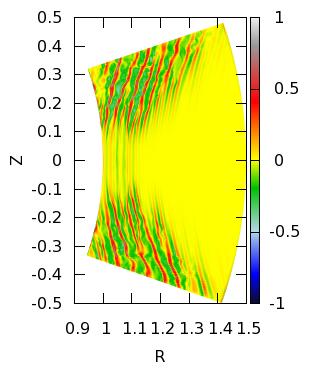}
\hspace{0mm}
\includegraphics[height=65mm]{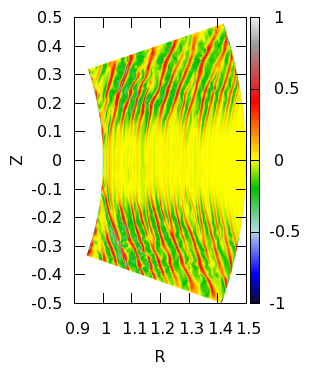}
\hspace{0mm}
\includegraphics[height=65mm]{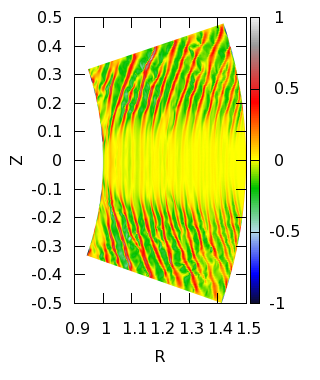}

\caption{Vorticity profile in the midplane and meridional plane after 
331, 555 and 694 orbital periods, with $h=0.1$ and $\tau=0.2$.}
\label{h=0.1}
\end{figure*}

The results of the simulation with $h=0.1$ and $\tau=0.2$ are shown in Fig.~\ref{h=0.1}.
As expected, the growth time of the instability is long in this case and the vorticity
perturbations that arise are of relatively low amplitude. The vortices that form are
observed to have aspect ratios $\chi \sim 10$, smooth non-turbulent cores,
and lifetimes that exceed the simulation run time of 695 orbits. The vertical heights of 
the vortices formed near the disc midplane again appear to extend approximately plus-and-minus 
one scaleheight about the midplane, as observed for the long-lived vortices formed
in the run with $h=0.2$ and $\tau=0.5$.

\begin{figure*}
\includegraphics[height=50mm]{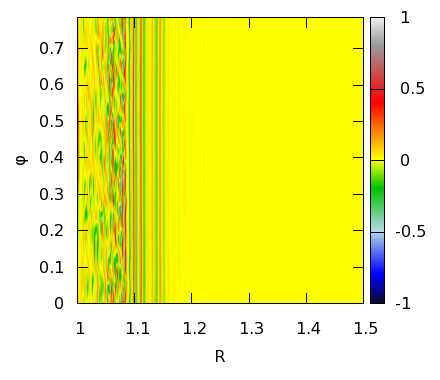}
\includegraphics[height=50mm]{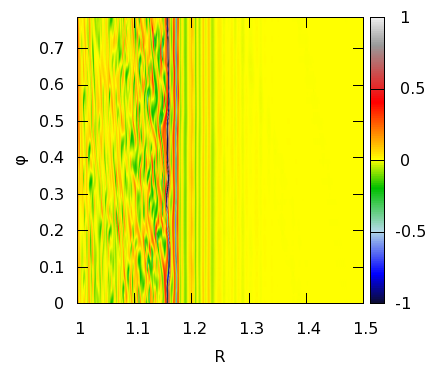}
\includegraphics[height=50mm]{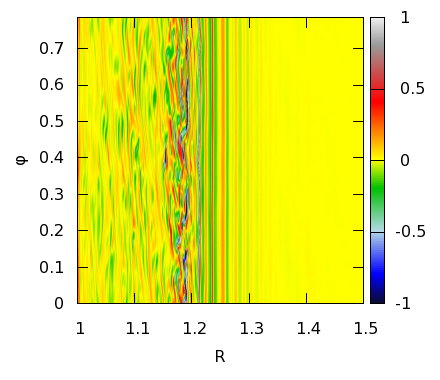}\\
\includegraphics[height=50mm]{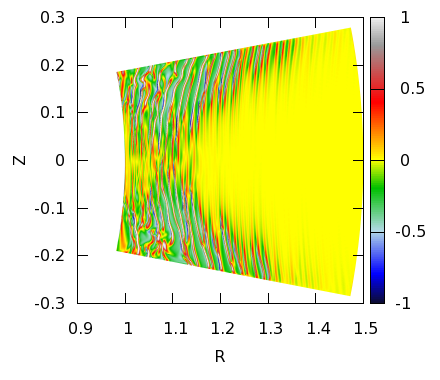}
\includegraphics[height=50mm]{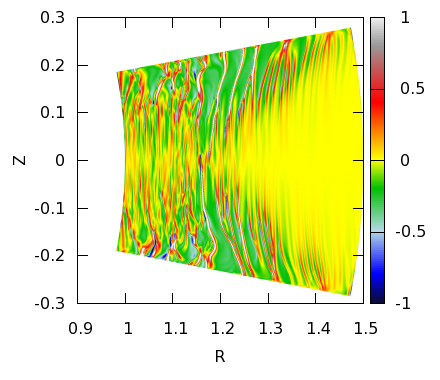}
\includegraphics[height=50mm]{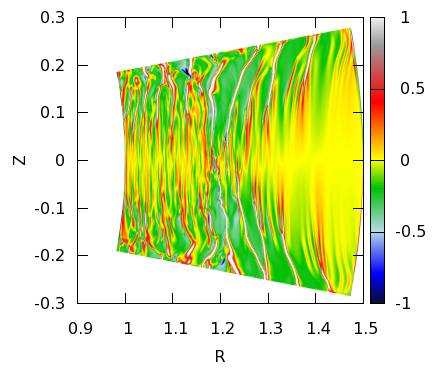}

\caption{Vorticity profile in the midplane and meridional plane after 
100, 120 and 130 orbital periods, with $h=0.05$ and $\tau=0.01$.}
\label{h=0.05}
\end{figure*}

The results for the simulation with $h=0.05$ and $\tau=0.01$ are shown in 
Fig.~\ref{h=0.05}. The short growth time of the VSI leads to the formation of
vortices with small aspect ratios $\chi \sim 1.5$ with lifetimes of the order
of one orbital period.

\subsection{Positive entropy gradient}
One of the original goals of this paper was to see if the VSI could produce strong
enough vorticity perturbations to trigger the SBI. In their study of vortex growth
in astrophysical discs, \citet{Petersen2007a} showed that vortices can form in the 
presence of applied cooling, whether or not the radial entropy stratification was stable 
or unstable. They also showed that vortices are amplified and maintained over long time 
scales when the stratification is unstable, but decay when the stratification is stable.
They found that vortex amplification is more efficient for short cooling time, but vortex 
formation through the SBI is more efficient if the cooling time is long.

In the previous simulations, we observed the formation of vortices even for cooling 
times as short as $\tau=0.01$. As discussed earlier in the paper, vortices appear 
to form from the axisymmetric perturbations to the vorticity that are generated 
by the VSI, indicating that vortex formation occurs as a secondary process driven 
by the RWI acting on the perturbations generated by the VSI. This suggests that the
SBI is not responsible for vortex formation, but may nonetheless play a role in
amplifying and maintaining long lived vortices in the flow if appropriate conditions
are present in the disc.

\begin{figure*}
\includegraphics[height=50mm]{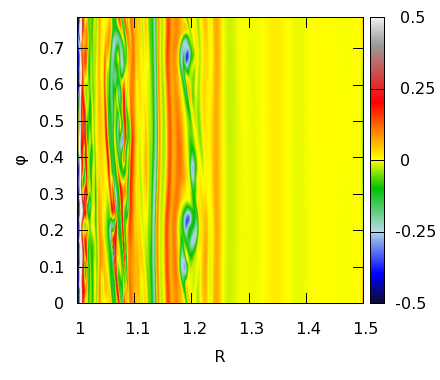}
\includegraphics[height=50mm]{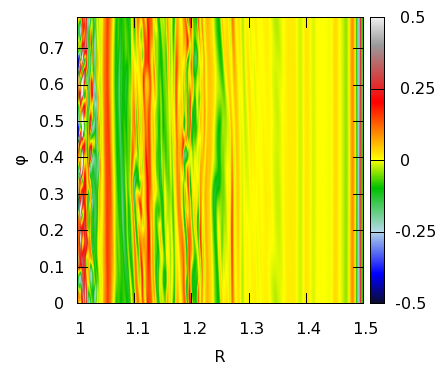}
\includegraphics[height=50mm]{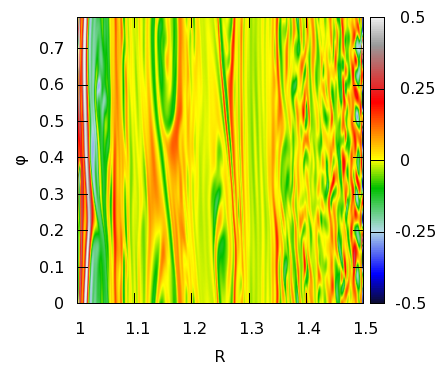}\\
\includegraphics[height=60mm]{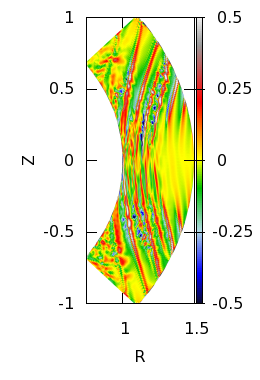}
\hspace{20mm}
\includegraphics[height=60mm]{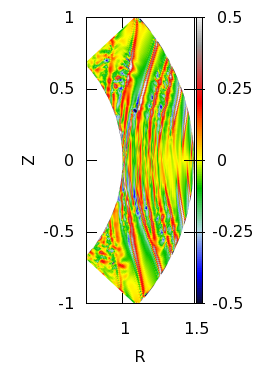}
\hspace{20mm}
\includegraphics[height=60mm]{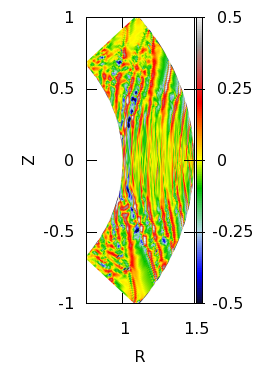}

\caption{Vorticity profile in the midplane and meridional plane after 
207, 237 and 295 orbital periods, in a disc with a positive entropy gradient in the midplane.}
\label{peg}
\end{figure*}

In an attempt to clarify this point, we present two simulations that have either a 
positive or negative midplane entropy gradient, respectively. The model with a positive 
entropy gradient has disc parameters $h=0.2$, $q=-1$, $p=-3$ and $\tau=0.1$
(note the shallower temperature profile). 
Fig. \ref{peg} shows the evolution of the vorticity in this simulation.
The results of this run are very similar to those described earlier in the
paper (which all had negative entropy gradients in the midplane): the VSI develops 
axisymmetrically, and vortices are formed from the band of perturbed vorticity that
is created by the VSI. We find that these vortices have quite large aspect ratios
($\chi \sim 8$), but the vortex lifetimes measured in the simulation are typically
in the range 15--20 orbital periods, considerably shorter than those measured
in the runs discussed earlier in this paper where large aspect ratio vortices 
were formed. 

The simulation with a negative midplane entropy gradient that was run to compare with 
the previous run with positive entropy gradient had identical parameters 
($h=0.2$, $q=-1$, $\tau=0.1$) except
for the midplane density power law which was set to $p=-2$. \citet{Nelson2013} 
showed that the density power law index makes essentially no difference to the
growth rate of the VSI, so we expect the VSI to develop in much the same way during
the early stages of these two runs. This is exactly what we observe, and in fact the
long term outcome of the simulation is very similar to the one with positive entropy
gradient: vortices with aspect ratios $\chi \sim 8$ are formed, and these live for
approximately 15 orbital periods before dissolving into that background flow.
This result indicates that for this particular set of disc parameters, the SBI has
essentially no influence on the formation or amplification of vortices.

The results obtained in this section are somewhat puzzling as it is to be expected
that the SBI should cause the amplification of the vortices that are formed by
the VSI. Furthermore, the run described earlier in Section~\ref{sect:cool-time}
with $h=0.2$, $p=-1.5$, $q=-2$ and $\tau=0.5$ showed evidence that vortices formed 
by the VSI were maintained against decay by the action of the SBI. In order to explore 
this issue further, we allowed
the simulation with negative entropy gradient described earlier in this section
to evolve until it had reached a saturated state, and we then restarted it with 
longer cooling times (both $\tau=0.5$ and $\tau=1$). In both cases we observed that
the vortices decayed and the VSI stopped operating, such that the disc evolved
towards a laminar state. This suggests that either the cooling time scale needs to
be fine-tuned in order to observe the SBI in this model, or perhaps a more likely
explanation is that the disc model with $h=0.2$, $p=-1.5$, $q=-2$ and
$\tau=0.5$ has a steeper entropy gradient which allows the SBI to operate more
strongly in that disc, whereas in a disc with weaker entropy gradient the SBI
at best operates very weakly and has little influence on the formation and
lifetimes of vortices.

\subsection{Disc with more realistic density and temperature profiles}
In this section, we investigate what might be thought of as a disc 
with a more realistic density and temperature profile than the
previous models. We adopt $p=-1.5$, $q=-1$, $h=0.05$ and $\tau=0.01$. 

In the previous simulations the parameters of the disc were chosen to
favour rapid growth of the instability. The temperature power law was set 
to $q=-2$ because the growth rate of the VSI is proportional to $|q|$,
and it gives a strong negative entropy gradient in the midplane.
The scaleheight was set to $h=0.2$ to obtain faster growth of the 
VSI and to increase the entropy gradient. These parameters also
lead to a disc that is unstable to the VSI with a cooling time 
that is long enough to also allow the development of the SBI.
Our previous simulations, however, show that vortices more likely
form due to the RWI rather than the SBI, so a strong entropy gradient
and long cooling time is not necessary to form vortices.

\begin{figure*}
\includegraphics[height=50mm]{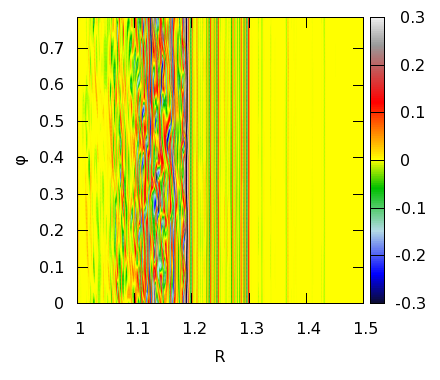}
\includegraphics[height=50mm]{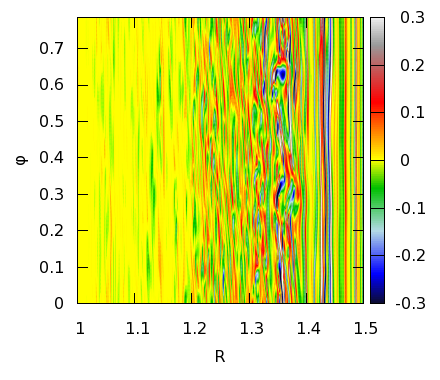}
\includegraphics[height=50mm]{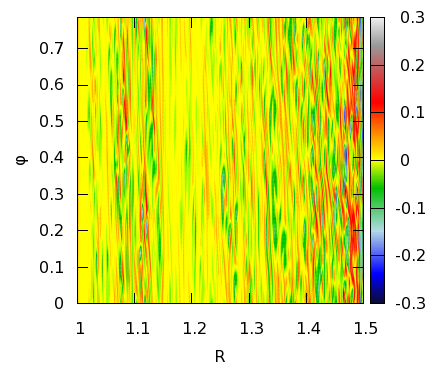}\\
\includegraphics[height=55mm]{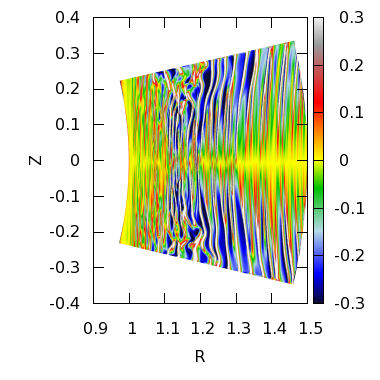}
\includegraphics[height=55mm]{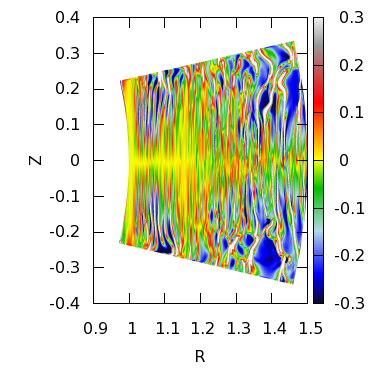}
\includegraphics[height=55mm]{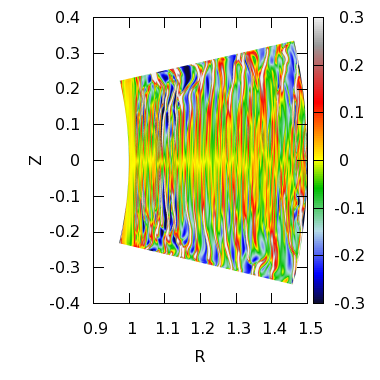}\\

\caption{Vorticity profile in the midplane and meridional plane after 
198, 254 and 291 orbital periods, with $H/R=0.05$, $q=-1$ and $\tau=0.01$.}
\label{rd}
\end{figure*}

Fig. \ref{rd} shows the result of the simulation with $h=0.05$, $p=-1.5$, $q=-1$ 
and $\tau=0.01$. As for the previous simulations with a thin disc, the radial 
resolution is doubled from $N_r=500$ to $N_r=1000$.
As usual, the VSI develops axisymmetrically. When the rings of vorticity destabilize, 
they give rise to vortices that survive for approximately two orbits and the disc
displays small scale turbulence. This turbulence tends to decay and after a few orbits
the disc relaminarizes. 
Then new VSI modes start to grow, and the process resumes, resulting in intermittent 
behaviour in which turbulence and small scale vortices regularly appear and disappear.

\subsection{Transport properties}

We now investigate the accretion rate resulting from the turbulence generated by the VSI.
We defined the Reynolds stress through:

\begin{equation}
T_r(r,\theta)={1\over \phi_{max}} \int_0^{\phi_{max}} \rho \delta v_r \delta v_\phi d\phi
\end{equation}
where $\delta v_r$ and $\delta v_\phi$ correspond to the local radial and azimuthal 
velocity fluctuations:
\begin{eqnarray}
\delta v_r & = & v_r-\langle v_r\rangle_\phi  \nonumber \\
\delta v_\phi& = & v_\phi-\langle v_\phi\rangle_\phi.
\end{eqnarray}
The local stress parameter $\alpha$ is obtained by dividing the Reynolds stress 
by the density-weighted mean pressure :
\begin{equation}
\bar{P}(r)={\int\int \rho P \sin(\theta)d\phi d\theta \over \int\int \rho \sin(\theta)d\phi d\theta}.
\end{equation}
 
\begin{table}
\begin{center}
\begin{tabular*}{0.45\textwidth}{@{\extracolsep{\fill}}  c c c c c}
\hline
$\tau$ & $h$ & $q$ & $p$ & $\alpha$ \\
\hline
0.05 & 0.2 & -2 & -1.5 & $2.65 \times 10^{-4}$\\
0.1 & 0.2 & -2 & -1.5 & $5.33  \times 10^{-5}$\\
0.5 & 0.2 & -2 & -1.5 & $2.8  \times 10^{-5}$\\
0.2 & 0.1 & -2 & -1.5 & $2.8  \times 10^{-6}$\\
0.01 & 0.05 & -2 & -1.5 & \\
0.1 & 0.2 & -1 & -3 & $2.8  \times 10^{-6}$\\
0.1 & 0.2 & -1 & -2 & $3.5  \times 10^{-6}$\\
0.01 & 0.05 & -1 & -1.5 & \\
\hline
\end{tabular*}
\end{center}
\caption{\label{Table_alpha}
The first column gives the cooling time, the second column the disc aspect ratio, 
the third and fourth columns the temperature and density power law indices, and
the last column gives the effective viscous stress parameter $\alpha$ obtained in
the simulations.}
\end{table}

The volume- and time-averaged $\alpha$ value for each simulation is listed in
Table~\ref{Table_alpha}, where the time average was taken over the last 20 orbits
of each run. There is a general tendency for the Reynolds stress to decrease as
the cooling time increases, due to the velocity fluctuations generated by the
VSI decreasing with increasing cooling time.
We also note that the simulation with $h=0.1$, which sustains smooth long-lived
vortices in a relatively quiet background flow without strong turbulence has
the lowest $\alpha$ value.

\begin{figure*}
\includegraphics[width=84mm]{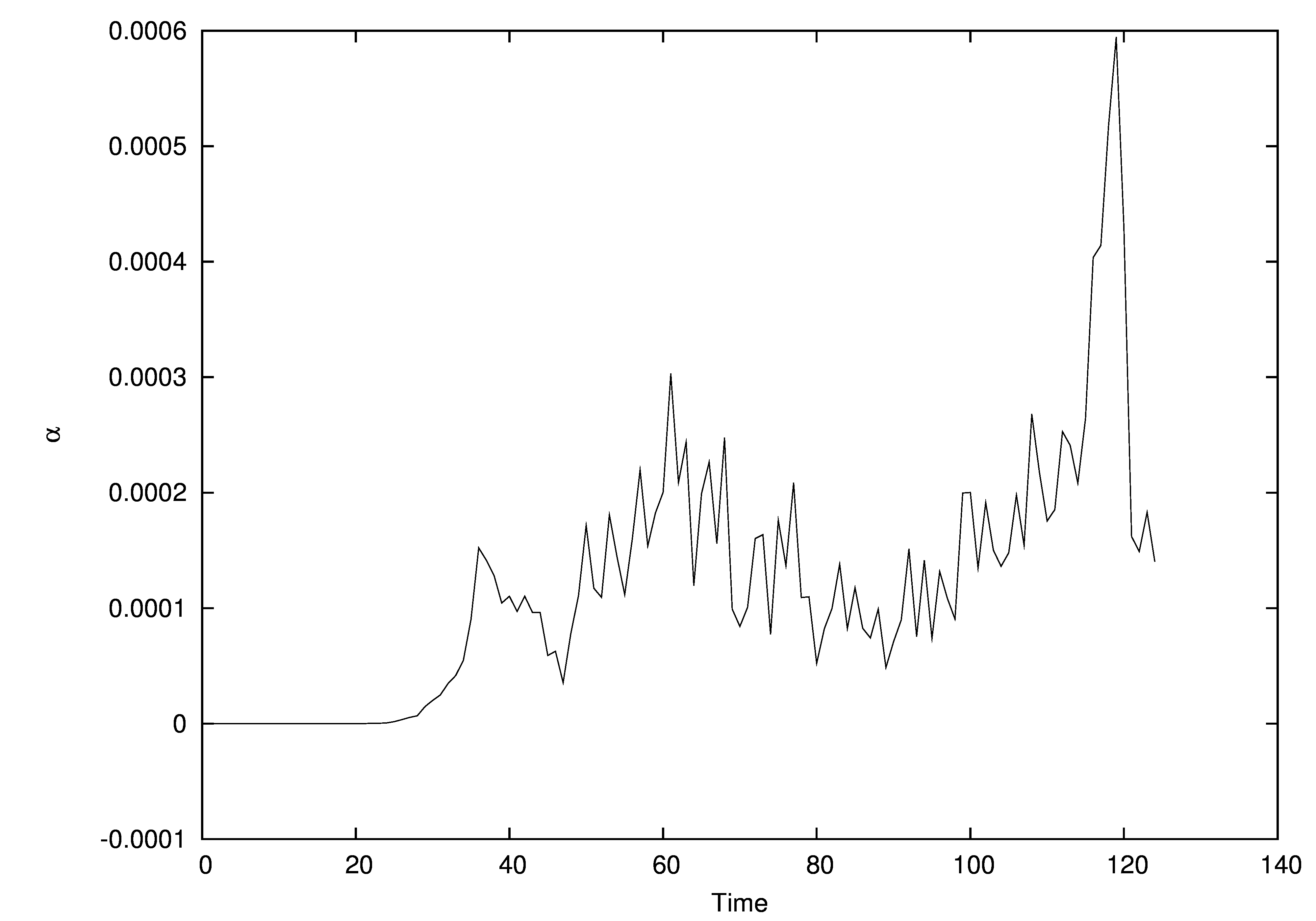}
\includegraphics[width=84mm]{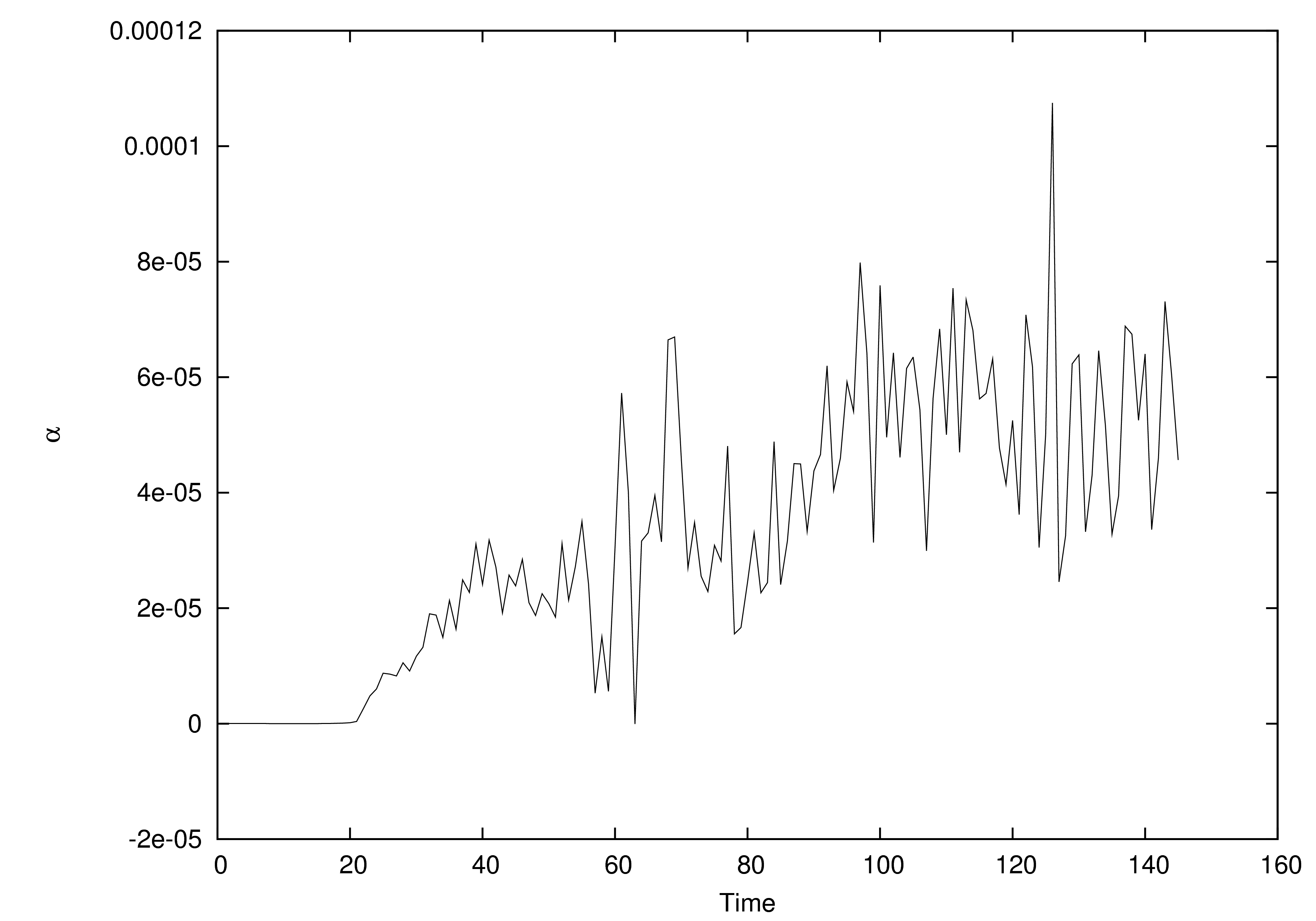} \\
\includegraphics[width=84mm]{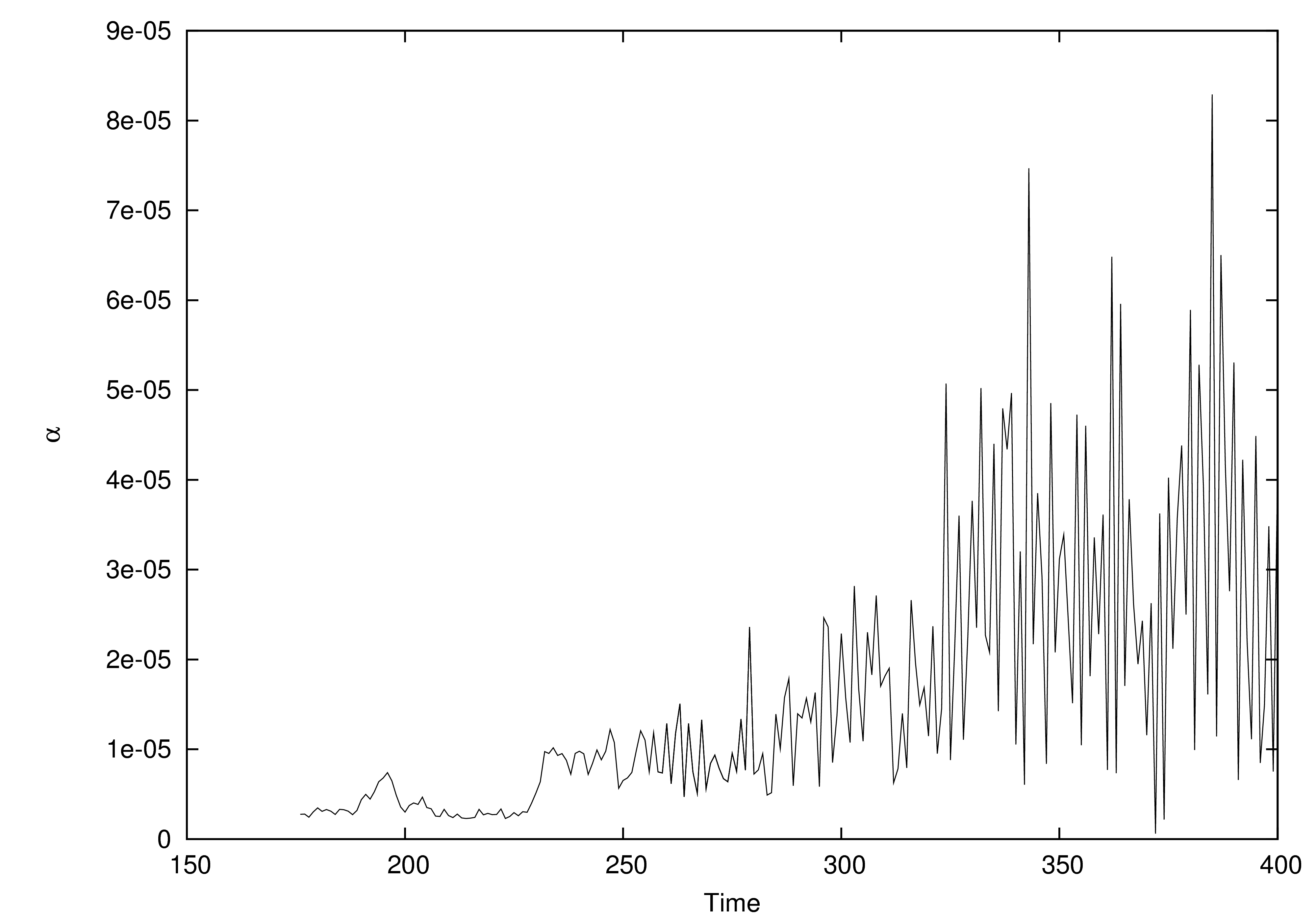}
\includegraphics[width=84mm]{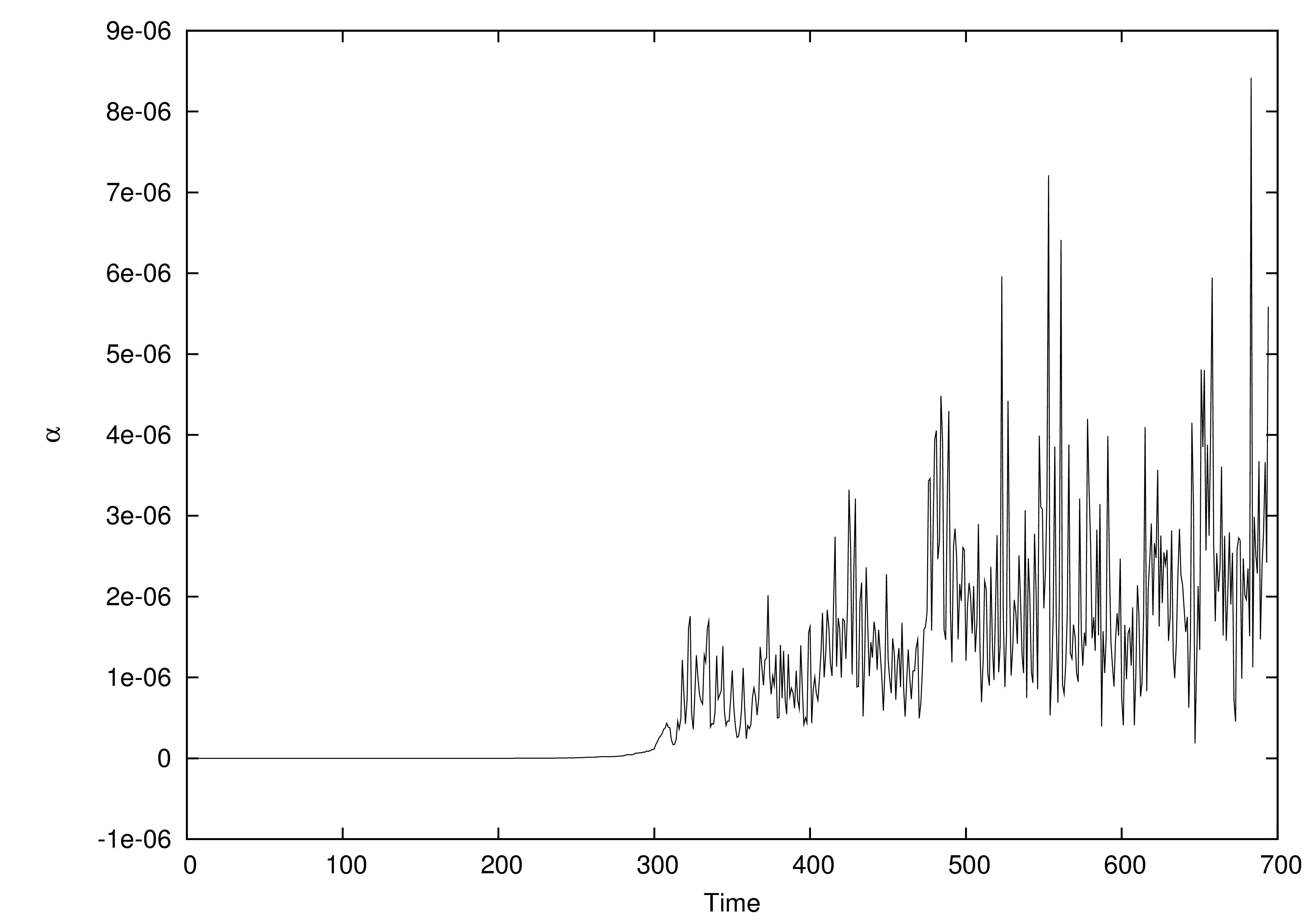}
\caption{Mean $\alpha$ value for :  a disc with a $h/r=0.2$ and $\tau=0.05$ (top left), a disc with a $h/r=0.2$ and $\tau=0.1$ (top right), a disc with a $h/r=0.2$ and $\tau=0.5$ (bottom left) and a disc with a $h/r=0.1$ and $\tau=0.2$ (bottom right).}
\label{alphas}
\end{figure*}

We also plot the temporal evolution of the mean $\alpha$ value for each of the
simulations with $h=0.2$, $q=-2$ and $p=-1.5$ in Fig.~\ref{alphas}.
The simulation with $\tau=0.05$ is shown in the top left panel, which displays
a peak value of $\alpha=6 \times 10^{-4}$. The simulation with $\tau=0.1$ is plotted 
in the top right panel, and shows a peak value of $\alpha \sim 10^{-4}$.
The evolution of $\alpha$ for the run with $\tau=0.5$ is shown in 
the bottom left panel and displays a peak value of $\alpha \sim 8 \times 10^{-5}$,
with typical values $\alpha \sim 5 \times 10^{-5}$. Finally, the evolution of
$\alpha$ for the run with $h=0.1$, $q=-2$ and $\tau=0.2$ is shown in the
bottom right panel of Fig.~\ref{alphas}. Here we see that the smooth flow
associated with this run containing long lived and high elongated vortices
produces a small Reynolds stress that has a stress parameter that has typically
values of $\alpha \sim 2 \times 10^{-6}$ during the simulation.

\section{Discussion and Conclusions}
\label{sec:discussion}
We have presented a series of customized and idealized simulations to
examine the nonlinear saturated state of astrophysical discs undergoing the
VSI, with the focus being on the formation and evolution
of vortices. The results of these simulations may be summarized as follows.
\begin{itemize}
\item We find that the VSI readily leads to the formation of vortices in a broad range of
disc models.
\item 
{{The formation mechanism of the vortices is the generation of axisymmetric 
vorticity perturbations by the VSI that destabilize into discrete vortices
through the RWI.  Specifically speaking, the VSI generates pressure perturbations
that induces the formation of narrow azimuthal jets.  These jets are characterized
by azimuthally symmetric strips of positive/negative vertical vorticity.  As a given jet grows in amplitude,
that side of it with the negative vorticity
anomaly becomes unstable to the RWI
and rolls-up into localized
vortices while the positive anomaly remains intact since it itself is not unstable to the RWI.  
As a given simulation
progresses and the VSI reaches out to larger radii, more and more unsteady vortices are produced
eventually spreading across the totality of the computational domain.}}
\item For steep temperature profiles and short cooling times ($h=0.2$, $q=-2$, $\tau \le 0.1$ orbits)
we observe that vortices form with small aspect ratios ($\chi \le 2$) and extend in height over 
length scales much smaller than the local scaleheight ($\sim 0.1 H$). These vortices
have short lifetimes of only a few orbital periods before dissolving into the background flow,
after which new generations of vortices form and dissolve in a repetitive cycle.
\item For longer cooling times and steep temperature (and entropy) profiles
($h=0.2$, $q=-2$, $\tau=0.5$ orbits) we observe the formation of elongated vortices ($\chi \sim 6$)
that live for hundreds of orbits (longer than the simulation run times). These vortices
live with a noticeably turbulent core, indicating that there is a balance between
the elliptical instability try to destroy the vortex and the SBI
maintaining the vortex over long time-scales. This suggests that the SBI can play an
important role in maintaining long lived vortices when the negative entropy profile is 
sufficiently steep. 
\item Similar behaviour is observed for thinner discs with $h=0.1$ and 0.05 when the
temperature profile has a steep power law index $q=-2$.
\item Changing the temperature power law index to $q=-1$ results in vortices that do not
live for more than a few tens of orbits at most. Simulations performed with either weakly
positive or negative entropy profiles produce very similar results, indicating that the influence
of the SBI is significantly diminished when the entropy profile is weaker than that obtained
with $q=-2$. 
\item Analysis of the volume averaged Reynolds stresses associated with the VSI
shows that the efficiency of angular momentum transport depends strongly on 
the thermal relaxation time scale. In line with our observation that the VSI generates 
stronger velocity fluctuations for short cooling times, we obtain an effective Shakura-Sunyaev 
$\alpha$ value \citep{Shakura73} of 
$\alpha_{\rm SS} \sim 2 \times 10^{-4}$ when $\tau=0.05$ for a model with $q=-2$ and $h=0.2$.
When $\tau=0.5$ we obtain $\alpha_{\rm SS} \sim 3 \times 10^{-5}$. These values 
should be contrasted with the value $\alpha_{\rm SS} \sim 10^{-3}$ obtained by 
\citet{Nelson2013} for a locally isothermal disc model, indicating that the inclusion
of a finite cooling time scale significantly reduces the strength of the VSI.
This point has also been noted by \citet{StollKley2014}, who performed simulations
with radiation transport and obtained $\alpha_{\rm SS} \sim 10^{-4}$.
\end{itemize}

The requirement of short cooling times for the VSI to operate indicates that
it is most likely to be present in the outer regions of protoplanetary
discs, between $\sim$ 10 and 50 au \citep{Nelson2013, Umurhan2013, LinYoudin2015},
although the short cooling times expected at higher disc latitudes may allow it
to also operate there at smaller stellocentric radii \citep{LinYoudin2015}.
The potential role of vortices in protoplanetary discs as sites for the
trapping of solids and growing planets has been well documented \citep{BargeSommeria1995},
and it is of interest to address the question of whether or not the vortices 
formed by the VSI are likely to play an important role in planet building.
The main difficulty with the VSI vortices trapping solids
is their apparent short lifetimes in the absence of a sufficiently steep
entropy profile. To assess the expected midplane entropy profiles in protoplanetary discs,
we assume that the disc surface density profile may be 
written as $\Sigma(R) = \Sigma_0 R^{\delta}$, and that the requirement for
a negative entropy gradient to exist locally at the midplane
(such that the SBI can play some role in extending vortex lifetimes) can be 
written as $q+p(1-\gamma) < 0$. Then we can examine which values of $\delta$ 
arise for reasonable values of the temperature power-law index such that the 
radial Brunt-Vaisala frequency obeys $N_R^2 < 0$. We note that the requirement 
for $N_R^2 <0$ can be written more conveniently as $p > q/(\gamma-1)$.

The minimum mass solar nebula model has $T(R) = 280 {\rm K} (R/ {\rm au})^{-1/2}$
\citep{Hayashi1981}. The more sophisticated passively irradiated disc model of
\citet{ChiangGoldreich1997} has $T(R) = 120 K (R/{\rm au})^{-3/7}$, i.e. the 
temperature profile is slightly shallower than the $q=-1/2$ value obtained from 
the Hayashi model. Observational constraints in the outer regions of discs 
indicate that $-0.7 \le q \le -0.4$ \citep{AndrewsWilliams2005}, covering the 
theoretically expected range. Adopting $q=-1/2$ and $\gamma=7/5$, we require 
$p > -5/4$ for a negative entropy gradient at the midplane. Writing 
$\Sigma \sim H \rho_{\rm mid}$, where $H$ is the scaleheight
and $\rho_{\rm mid}$ is the midplane density, and using $p=-5/4$ and
$H(R) \propto R^{(3+q)}/2$ (from the definition $H=c_{\rm s}/v_{\rm k}$),
we see that the power law index for the surface density $\delta=0$ for this model.
In other words, a disc that has $q=-1/2$ and \emph{just} supports a large scale 
negative entropy gradient in its outer regions must have a flat or outwardly 
increasing surface density profile. Observations suggest that surface density
power law values in the outer regions of discs are typically $\delta \sim -1$
\citep{WilliamsCieza2011}, such that the entropy profile is increasing and
not decreasing outwards. Our simulations indicate that just having
a modestly decreasing entropy profile is unlikely to be sufficient for maintaining
long lived vortices, and that the entropy gradient needs to be reasonably
steep. The above discussion suggests that the globally inferred temperature and 
surface density profiles in the outer regions of protoplanetary discs are unlikely 
to support the existence of long-lived vortices maintained by the SBI. We note,
however, that we cannot rule out the possibility that shadowing or other effects
may increase the temperature gradient locally such that vortices can exist as
long-lived structures around the midplane.  
{{We note recent work
 studying particle
growth in evolving protoplanetary discs
\citep{Estrada2015}
wherein 
discs with complex temperature and opacity profiles are found ones
in which the VSI is likely to be active in several localized radial
sections, including places near to various ice lines.}}
\par
We conclude that although the VSI may operate in the outer regions of discs,
and produce vortices, these are likely to be relatively short lived structures
that at best play a moderate role in trapping solids and assisting in the
building of planets. The presence of a quasi-turbulent flow arising from the VSI
may indeed provide a source of stirring that could act to hinder the formation
of planetesimals and planets \citep{Gressel2011,Gressel2012} or perhaps
restrict it into places in the disc where the VSI is either weak or entirely absent.

Having presented a set of idealized models with simple cooling prescriptions
to examine the evolution of the VSI in protoplanetary discs, the next step is
to undertake multidimensional radiation-hydrodynamic simulations with realistic
opacities to examine the nonlinear outcome of the VSI in more realistic disc models.
These calculations will be the subject of future publications.

\section*{Acknowledgements}
This work used the DiRAC Complexity system, operated by the University of Leicester
IT Services, which forms part of the STFC DiRAC HPC Facility (www.dirac.ac.uk).
The equipment is funded by BIS National E-Infrastructure capital grant ST/K000373/1
and STFC Operations grant ST/K0003259/1. DiRAC is part of the national E-Infrastructure.

\bibliographystyle{mn2e}
\bibliography{vsi}

\end{document}